\documentclass[preprint,showpacs,preprintnumbers,amsmath,amssymb]{revtex4}
\usepackage{graphicx}
\usepackage{dcolumn}
\usepackage{bm}
\usepackage{amssymb}
\usepackage{amsmath}
\usepackage{latexsym}

\begin{document}

\title{Non-Markovian feature of the classical Hall effect
}
\author{
I.B. Abdurakhmanov$^{1}$,
G.G. Adamian$^{2}$,  N.V. Antonenko$^{2}$, and Z. Kanokov$^{2,3,4}$,
}
\affiliation{
$^{1}$Curtin Institute for Computation, Department of Physics and Astronomy, Curtin University,  Perth, WA 6845, Australia\\
$^{2}$Joint Institute for
Nuclear Research, 141980 Dubna, Russia\\
$^{3}$National University, 700174 Tashkent,
Uzbekistan\\
$^{4}$Institute of Nuclear Physics,
702132 Tashkent, Uzbekistan
}

\date{\today}

\begin{abstract}
 The classical Hall effect
 resulting from  the impact of external magnetic
and electric fields on the non-Markovian dynamics of
charge carriers is studied.
The dependence of the tangent of the Hall angle on the magnetic field
is derived and compared with the experimental data for Zn.
The method is proposed to determine  experimentally the memory time in a system.
\end{abstract}

\maketitle
\section{ Introduction}
The behavior of solid matter under the influence of impact fields is one of the interesting topics
in solid-state physics \cite{Von1,Von2,Von3,new,new2}. The external field may be an electric field, a magnetic field,
an optical signal, or a temperature gradient. Such external fields modify the distribution
of   internal energy which in turn modifies or alters the electronic properties, such as
the carrier concentration or the carrier mobility. Besides the carrier mobility, the electric
current is also affected by  magnetic field  which deflects its direction. Modeling the
electric current implies the determination of the time-dependence of the number of electrons with the given momentum at certain
location. The equations of motion for it can be obtained by using the quantum
Langevin approach or density matrix formalism
which is widely applied to find the effects of fluctuations and
dissipation in macroscopic systems  \cite{Kampen1,Kampen2,Kampen3,Kampen4,Kampen5,Kampen6,LEG,DM,Katia,Isar,M1100}. The Langevin method in the kinetic theory
significantly simplifies the calculation of nonequilibrium
quantum and thermal fluctuations and provides a clear picture
of the dynamics of the process \cite{Katia,M1100,Hu1,Hu2,Ford,Ford1,In,M110,Kanokov,Lac9,Lac10,Lac11,Lac12,Lac13,Lac14,Lac15,Lac16,Lac17}.

The aim of the present work is the
treatment of the classic Hall effect in the external constant magnetic and electric fields
beyond the Markov approximation (instantaneus dissipation and delta-correlated fluctuations)
and the weak-coupling limit. We generalize the Langevin formalism,
which has been developed for non-Markovian noise in Refs.~\cite{Hu2,Kanokov},
by including the external fields.
The basic idea of our model is the following:
we consider the center of mass of
the charge carriers with the positive charge $e=|e|$ as a quantum particle
  coupled to the environment (heat bath) through
particle-phonon interactions. In the   solution of the second order
Heisenberg equations for the heat bath degrees of freedom, the
generalized non-Markovian Langevin equations of a quantum particle
is  directly obtained. At the same time,
the memory function, which contains all the information concerning the effect
of  heat bath on the transport properties of the quantum particle, is also
obtained without  any approximations for the
particle-phonon  interaction.

The paper is organized as follows.
In Sec.~II, we give the Hamiltonian of
the system and solve the
generalized non-Markovian Langevin equations for a quantum particle.
In Sec. III, we   consider linear coupling in the coordinate between the heat bath and
 quantum particle which describes the center
of mass of the charge carriers.
The effects of heat-bath and external  constant electric
magnetic field on the dynamics of quantum particle
are fully studied. The analytical expressions for
 macroscopically observable values
are worked out in the context of the  model. The classical Hall effect   is considered.
The cross component of electric field (which is absent at the initial time) generating
by the   magnetic field and heat bath is investigated
at   different system conditions.
The   developed model is used in Sec. IV   to  describe   the experimental
data on the classical Hall effect
in   Zn. The calculated
  results are shown to be in a good agreement   with the experiment.

\section{ Non-Markovian Langevin equations with external magnetic and electric fields}
\subsection{Derivation of quantum Langevin equations}
Let us consider the two-dimensional  motion of   quantum charged particle in the presence
of heat bath and
 external constant electric  ${\bf E}=(E_x,0,0)$ and magnetic fields ${\bf B}=(0,0,B)$.
 The total
Hamiltonian of this system is
\begin{eqnarray}
H=H_c+H_b+H_{cb}.
\label{equ1}
\end{eqnarray}
The Hamiltonian $H_c$   describes the collective subsystem (quantum particle) with
effective mass tensor and charge $e=|e|$  in electric and magnetic fields:
\begin{eqnarray}
H_c
= \frac{1}{2 m_x}(p_x-e A_x(x,y))^2+\frac{1}{2 m_y}(p_y-e
A_y(x,y))^2+e E_x x= \frac{\pi_x^2}{2m_x}+\frac{\pi_y^2}{2m_y}+e
E_x x.
\label{equ2}
\end{eqnarray}
Here,  $m_x$ and $m_y$ are the components of the effective mass
tensor, ${\bf R}=(x,y,0)$ and
${\bf p}=(p_x,p_y,0)$ are the coordinate and canonically conjugated momentum,
respectively, ${\bf A}=(-\frac{1}{2}y B,\frac{1}{2}x B,0)$ is
 the vector potential of the magnetic field,
and the electric
field  $E_x$ acts in   $x$ direction. For simplicity, in Eq.~(\ref{equ2})
we introduce  the notations
$$\pi_x=p_x+\frac{1}{2}m_x\omega_{cx}y,\hspace{.3in}
\pi_y=p_y-\frac{1}{2}m_y\omega_{cy}x$$ with  frequencies
 $\omega_{cx}=\frac{e B}{m_x }$ and $\omega_{cy}=\frac{e
B}{m_y }$.
The cyclotron frequency is $\omega_c=\sqrt{\omega_{cx} \omega_{cy}}=\frac{e
B}{\sqrt{m_x m_y}}$.

The second term in Eq.~(\ref{equ1}) represents the Hamiltonian of the phonon
heat bath,
\begin{eqnarray}
H_b=\sum_{\nu}^{}\hbar\omega_\nu b_\nu^+b_\nu,
\label{equ3}
\end{eqnarray}
where $b_\nu^+$ and $b_\nu$ are the phonon creation and
annihilation operators of the heat bath. The coupling between the
heat bath and   collective subsystem is described by
\begin{eqnarray}
H_{cb}=\sum_{\nu}^{}V_\nu({\bf R})(b_\nu^+ +
b_\nu)+\sum_{\nu}^{}\frac{1}{\hbar\omega_\nu}V_\nu^2({\bf R}).
\label{equ4}
\end{eqnarray}
The first term of Eq. (\ref{equ4}) corresponds to the exchange of   energy between
the collective subsystem and heat bath. We introduce the
counterterm (second term) in $H_{cb}$ in order to
compensate the coupling-induced renormalization of the potential.
Naturally, it can   be always splitted off from $e E_x x$ in
Eq.~(\ref{equ2}).
In general case, $V_\nu({\bf R})$ depends on a strength of
magnetic field and the impact of $\bf B$ is entered into the dissipative kernels
and random forces.

The equations of motion are
\begin{eqnarray}
\dot{x}(t)&=&\frac{i}{\hbar}[H,x]=\frac{\pi_x(t)}{m_x},\hspace{.3in}
\dot {y}(t)=\frac{i}{\hbar}[H,y]=\frac{\pi_y(t)}{m_y},\nonumber\\
\dot\pi_{x}(t)&=&\frac{i}{\hbar}[H,\pi_x]=\pi_{y}(t)\omega_{cy}-e
E_{x} -\sum_{\nu}^{}V'_{\nu,x}({\bf R})(b_\nu^+ + b_\nu)-
2\sum_{\nu}^{}\frac{V_{\nu}({\bf R})
V'_{\nu,x}({\bf R})}{\hbar\omega_{\nu}},\nonumber\\
\dot \pi_{y}(t)&=&\frac{i}{\hbar}[H,\pi_y]=-\pi_{x}(t)\omega_{cx}
-\sum_{\nu}^{}V'_{\nu,y}({\bf R})(b_\nu^+ + b_\nu)-
2\sum_{\nu}^{}\frac{V_{\nu}({\bf R})
V'_{\nu,y}({\bf R})}{\hbar\omega_{\nu}},
\label{equ5}
\end{eqnarray}
and
\begin{eqnarray}
\dot b_\nu^+(t)&=&\frac{i}{\hbar}[H,b_\nu^+]=
i\omega_\nu b_\nu^+(t) + \frac{i}{\hbar}V_{\nu}({\bf R}), \nonumber\\
\dot b_\nu(t)&=&\frac{i}{\hbar}[H,b_\nu]= -i\omega_\nu b_\nu(t)
- \frac{i}{\hbar}V_{\nu}({\bf R}).
\label{equ6}
\end{eqnarray}
The solution of Eqs.~(\ref{equ6}) are
\begin{eqnarray}
b_\nu^+(t)+b_\nu (t)&=&f^{+}_\nu (t)+f_\nu (t) -
\frac{2V_{\nu}({\bf R})}{\hbar\omega_\nu}  + \frac{2}{\hbar\omega_\nu}\int\limits_{0}^{t}d\tau
\dot V_{\nu}({\bf R}(\tau))\cos(\omega_\nu[t-\tau]),\nonumber\\
b_\nu^+(t)-b_\nu (t)&=&f^{+}_\nu (t)-f_\nu (t) +\frac{2i}{\hbar\omega_\nu}\int\limits_{0}^{t}d\tau
\dot V_{\nu}({\bf R}(\tau))\sin(\omega_\nu[t-\tau]),
\label{equ7}
\end{eqnarray}
where
\begin{eqnarray}
f_\nu(t)=[b_\nu(0)+\frac{1}{\hbar\omega_\nu}V_{\nu}({\bf R}(0))]e^{-i\omega_\nu t}. \nonumber
\end{eqnarray}
Substituting (\ref{equ7})
into (\ref{equ5}) and  eliminating the bath variables from
the equations of motion for the collective subsystem, we obtain a
set of nonlinear integro-differential stochastic dissipative
equations
\begin{eqnarray}
\dot {x}(t)&=&\frac{\pi_x(t)}{m_x}, \hspace{.3in}
\dot {y}(t)=\frac{\pi_y(t)}{m_y}, \nonumber\\
\dot \pi_{x}(t)&=&\pi_{y}(t)\omega_{cy} -e E_{x}-
 \frac{1}{2}\int\limits_{0}^{t}d\tau \{K_{xx}(t,\tau),\dot{x}(\tau)\}_{+}
 -\frac{1}{2}\int\limits_{0}^{t}d\tau \{K_{xy}(t,\tau),\dot{x}(\tau)\}_{+}+ F_{x}(t),\nonumber\\
\dot \pi_{y}(t)&=&-\pi_{x}(t)\omega_{cx} -
 \frac{1}{2}\int\limits_{0}^{t}d\tau \{K_{yy}(t,\tau),\dot{y}(\tau)\}_{+}
 -\frac{1}{2}\int\limits_{0}^{t}d\tau \{K_{yx}(t,\tau),\dot{y}(\tau)\}_{+}+ F_{y}(t).
\label{equ8}
\end{eqnarray}
The dissipative kernels and random forces in (\ref{equ8}) are
\begin{eqnarray}
K_{xx}(t,\tau)&=& \sum_{\nu}^{}\frac{1}{\hbar\omega_\nu}\{V'_{\nu,x}({\bf R}(t)),V'_{\nu,x}({\bf R}(\tau))\}_{+}\cos(\omega_\nu [t-\tau]), \nonumber\\
K_{xy}(t,\tau)&=& \sum_{\nu}^{}\frac{1}{\hbar\omega_\nu}\{V'_{\nu,x}({\bf R}(t)),V'_{\nu,y}({\bf R}(\tau))\}_{+}\cos(\omega_\nu [t-\tau]), \nonumber\\
K_{yx}(t,\tau)&=& \sum_{\nu}^{}\frac{1}{\hbar\omega_\nu}\{V'_{\nu,y}({\bf R}(t)),V'_{\nu,x}({\bf R}(\tau))\}_{+}\cos(\omega_\nu [t-\tau]), \nonumber\\
K_{yy}(t,\tau)&=&
\sum_{\nu}^{}\frac{1}{\hbar\omega_\nu}\{V'_{\nu,y}({\bf R}(t)),V'_{\nu,y}({\bf R}(\tau))\}_{+}\cos(\omega_\nu
[t-\tau])
\label{equ9}
\end{eqnarray}
and
\begin{eqnarray}
F_{x}(t)&=&\sum_{\nu}{}F_{x}^{\nu}(t)=-\sum_{\nu}{}V'_{\nu,x}({\bf R}(t))[f_{\nu}^{+}(t)+f_{\nu}(t)],\nonumber\\
F_{y}(t)&=&\sum_{\nu}{}F_{y}^{\nu}(t)=-\sum_{\nu}{}V'_{\nu,y}({\bf R}(t))[f_{\nu}^{+}(t)+f_{\nu}(t)],
\label{equ10}
\end{eqnarray}
respectively.
Here, we use the  notations: $V'_{\nu,x}=\partial V_{\nu}/\partial x$,
$V'_{\nu,y}=\partial V_{\nu}/\partial y$, and $\{Z_1,Z_2\}_+=Z_1Z_2+Z_2Z_1$.
Following the usual procedure in statistical mechanics,
we identify the  operators $F_{x}^{\nu}$ and $F_{y}^{\nu}$ as fluctuations because of the
uncertainty of the initial conditions for the bath operators. To
specify the statistical properties of the fluctuations, we
consider an ensemble of initial states in which the fluctuations
have the Gaussian distribution with zero average value
\begin{eqnarray}
\ll F^\nu_x (t)\gg = \ll F^\nu_y (t)\gg = 0.
\label{equ11}
\end{eqnarray}
Here, the symbol $\ll...\gg$ denotes the  average over the bath.
Bose-Einstein statistics of the bath are
\begin{eqnarray}
\ll f_{\nu}^{+}(t)f_{\nu'}^{+}(t')\gg &=&\ll f_{\nu}(t)f_{\nu'}(t')\gg=0, \nonumber\\
\ll f_{\nu}^{+}(t)f_{\nu'}(t')\gg &=&\delta_{\nu,\nu'}n_{\nu}e^{i\omega_\nu [t-t']},\nonumber\\
\ll f_{\nu}(t)f_{\nu'}^{+}(t')\gg
&=&\delta_{\nu,\nu'}(n_{\nu}+1)e^{-i\omega_\nu [t-t']},
\label{equ12}
\end{eqnarray}
where the occupation numbers $n_\nu=[\exp(\hbar\omega_\nu/T)-1]^{-1}$ for phonons depend  on temperature
$T$.

Using the properties   (\ref{equ10}) and (\ref{equ11}) of random forces,
we get the following
symmetrized correlation functions $\varphi_{kk'}^{\nu}(t,t')=\ll
F^{\nu}_k(t)F^{\nu}_{k'}(t')+F^{\nu}_{k'}(t')F^{\nu}_k(t)\gg$,
($k, k'=x, y$):
\begin{eqnarray}
\varphi_{xx}^\nu(t,t')&=&[2n_{\nu}+1]\{V'_{\nu,x}({\bf R}(t)),V'_{\nu,x}({\bf R}(t'))\}_{+}\cos(\omega_\nu [t-t']),\nonumber\\
\varphi_{yy}^\nu(t,t')&=&\varphi_{xx}^\nu(t,t')|_{{x\to y}},\nonumber\\
\varphi_{xy}^\nu(t,t')&=&[2n_{\nu}+1]\{V'_{\nu,x}({\bf R}(t)),V'_{\nu,y}({\bf R}(t'))\}_{+}\cos(\omega_\nu [t-t']),\nonumber\\
\varphi_{yx}^\nu(t,t')&=&\varphi_{xy}^\nu(t,t')|_{{x\to y}}.
\label{equ13}
\end{eqnarray}
The quantum fluctuation-dissipation relations read
\begin{eqnarray}
\sum_{\nu}^{}\varphi_{xx}^\nu(t,t')\frac{\tanh[\frac{\hbar\omega_\nu}{2T}]}{\hbar\omega_\nu}=K_{xx}(t,t'), \nonumber\\
\sum_{\nu}^{}\varphi_{yy}^\nu(t,t')\frac{\tanh[\frac{\hbar\omega_\nu}{2T}]}{\hbar\omega_\nu}=K_{yy}(t,t'),\nonumber\\
\sum_{\nu}^{}\varphi_{xy}^\nu(t,t')\frac{\tanh[\frac{\hbar\omega_\nu}{2T}]}{\hbar\omega_\nu}=K_{xy}(t,t'),\nonumber\\
\sum_{\nu}^{}\varphi_{yx}^\nu(t,t')\frac{\tanh[\frac{\hbar\omega_\nu}{2T}]}{\hbar\omega_\nu}=K_{yx}(t,t').
\label{equ14}
\end{eqnarray}
The validity of the fluctuation-dissipation relations means that we have properly identified the
dissipative terms in the non-Markovian dynamical equations of motion.
The quantum fluctuation-dissipation relations differ from the
classical ones and are reduced to them in the limit of high temperature.

\subsection{Derivation of non-stationary transport coefficients}
In order to solve the equations of motion (\ref{equ8}) for the collective
variables, we applied the Laplace transformation. It significantly
simplifies the solution of the problem. After the Laplace
transformation, the equations of motion take as
\begin{eqnarray}
x(s)s=x(0)+\frac{\pi_x(s)}{m_x}&,&\hspace{.2in}
y(s)s=y(0)+\frac{\pi_y(s)}{m_y}, \nonumber\\
\pi_{x}(s)s+\frac{\pi_{x}(s)}{m_x}(K_{xx}(s)+K_{xy}(s))&=&\pi_{x}(0)+\omega_{cy} \pi_{y}(s)-\frac{1}{s}e E_{x}+ F_{x}(s),\nonumber\\
\pi_{y}(s)s+\frac{\pi_{y}(s)}{m_y}(K_{yy}(s)+K_{yx}(w))&=&\pi_{y}(0)-\omega_{cx}
\pi_{x}(s)+ F_{y}(s).
\label{equ15}
\end{eqnarray}
Here,  $K_{xx}(s)$, $K_{yy}(s)$, $K_{xy}(s)$,   $K_{yx}(s)$ and $F_{x}(s)$, $F_{y}(s)$ are the Laplace transforms of the
dissipative kernels and random forces, respectively.
For the solution of this system of equations, one should find the roots of the
  determinant
\begin{eqnarray}
D=s(m_x m_y \omega_c^2+[K_{xx}(s)+K_{xy}(s)+m_x s][K_{yy}(s)+K_{yx}(s)+m_y s])=0.
\label{equ16}
\end{eqnarray}
The explicit solutions for the originals are
\begin{eqnarray}
x(t)&=&x(0)+ A_{1}(t) \pi_{x}(0)+A_{2}(t)\pi_{y}(0)-A_{3}(t)eE_{x}+I_x(t)+I'_x(t) ,\nonumber\\
y(t)&=&y(0)+ B_{1}(t) \pi_{y}(0)-B_{2}(t) \pi_{x}(0)+B_{3}(t)eE_{x}- I_y(t)+I'_y(t), \nonumber\\
\pi_x(t)&=& C_{1}(t) \pi_{x}(0)+C_{2}(t)\pi_{y}(0)-C_{3}(t)eE_{x}+I_{\pi_x}(t)+I'_{\pi_x}(t), \nonumber\\
\pi_y(t)&=&D_{1}(t)\pi_{y}(0)-D_{2}(t)\pi_{x}(0)+D_{3}(t)eE_{x}-I_{\pi_y}(t)+I'_{\pi_y}(t),
\label{equ17}
\end{eqnarray}
where
$$I_x(t)=\int_0^t A_{1}(\tau)F_{x}(t-\tau)d\tau, \hspace{0.3in}
I'_x(t)=\int_0^t A_{2}(\tau)F_{y}(t-\tau)d\tau,$$
$$I_y(t)=\int_0^t
B_{2}(\tau)F_{x}(t-\tau)d\tau, \hspace{0.3in}   I'_y(t)=\int_0^t
B_{1}(\tau)F_{y}(t-\tau)d\tau,$$
$$I_{\pi_x}(t)=\int_0^t
C_{1}(\tau)F_{x}(t-\tau)d\tau, \hspace{0.3in} I'_{\pi_x}(t)=\int_0^t
C_{2}(\tau)F_{y}(t-\tau)d\tau,$$
$$I_{\pi_y}(t)=\int_0^t
D_{2}(\tau)F_{x}(t-\tau)d\tau, \hspace{0.3in} I'_{\pi_y}(t)=\int_0^t
D_{1}(\tau)F_{y}(t-\tau)d\tau$$
and the following time-dependent
coefficients:
\begin{eqnarray}
A_{1}(t)&=&\hat{L}^{-1}\left[\frac{K_{yy}(s)+K_{yx}(s)+m_y s}{D}\right]=B_{1}(t)|_{x\leftrightarrow y}, \nonumber\\
A_{2}(t)&=&m_y \omega_{cy} \hat{L}^{-1}\left[\frac{1}{D}\right]=B_{2}(t)|_{x\leftrightarrow y},\nonumber\\
A_{3}(t)&=&\hat{L}^{-1}\left[\frac{K_{yy}(s)+K_{yx}(s)+m_y s}{s
D}\right],\hspace{0.3in}
B_{3}(t)=m_x \omega_{cx}\hat{L}^{-1}\left[\frac{1}{s D}\right],\nonumber\\
C_{1}(t)&=&m_x \hat{L}^{-1}\left[\frac{s(K_{yy}(s)+K_{yx}(s)+m_y s)}{D}\right]=D_{1}(t)|_{x\leftrightarrow y},\nonumber\\
C_{2}(t)&=&m_x m_y \omega_{cy} \hat{L}^{-1}\left[\frac{s}{D}\right]=D_{2}(t)|_{x\leftrightarrow y},\nonumber\\
C_{3}(t)&=&m_x \hat{L}^{-1}\left[\frac{K_{yy}(s)+K_{yx}(s)+m_y
s}{D}\right],\hspace{0.3in} D_{3}(t)=m_x m_y \omega_{cx}
\hat{L}^{-1}\left[\frac{1}{D}\right].
\label{equ18}
\end{eqnarray}
Here, $\hat{L}^{-1}$ denotes the inverse Laplace transformation.
The exact solutions of $x(t)$, $y(t)$, $\pi_x(t)$, and $\pi_y(t)$
in terms of roots $s_i$ are given by the residue theorem.

In order to determine the transport coefficients, we use Eqs.
(\ref{equ17}). Averaging them over the whole system and by
differentiating in $t$, we obtain the  system of equations for
the first moments:
\begin{eqnarray}
<\dot{x}(t)>&=&\frac{<\pi_x(t)>}{m_x},\hspace{.3in}
<\dot{y}(t)>=\frac{<\pi_y(t)>}{m_y},\nonumber\\
<\dot{\pi}_x(t)>&=&\tilde{\omega}_{cy}(t)<\pi_y(t)>-\lambda_{\pi_x}(t)<\pi_x(t)>-e \tilde{E}_{xx}(t),\nonumber\\
<\dot{\pi}_y(t)>&=&-\tilde{\omega}_{cx}(t)<\pi_x(t)>-\lambda_{\pi_y}(t)<\pi_y(t)>-e\tilde{E}_{xy}(t),
\label{equ19}
\end{eqnarray}
where the friction coefficients are
\begin{eqnarray}
\lambda_{\pi_{x}}(t)=-\frac{D_{1}(t)\dot{C}_{1}(t)+D_{2}(t)\dot{C}_{2}(t)}{C_{1}(t)D_{1}(t)+C_{2}(t)D_{2}(t)},
\nonumber\\
\lambda_{\pi_{y}}(t)=-\frac{C_{1}(t)\dot{D}_{1}(t)+C_{2}(t)\dot{D}_{2}(t)}{C_{1}(t)D_{1}(t)+C_{2}(t)D_{2}(t)},
\label{equ20}
\end{eqnarray}
and the renormalized cyclotron frequencies are given by
\begin{eqnarray}
\tilde{\omega}_{cx}(t)=\frac{D_{1}(t)\dot{D}_{2}(t)-D_{2}(t)\dot{D}_{1}(t)}{C_{1}(t)D_{1}(t)+C_{2}(t)D_{2}(t)},
\nonumber\\
\tilde{\omega}_{cy}(t)=\frac{C_{1}(t)\dot{C}_{2}(t)-C_{2}(t)\dot{C}_{1}(t)}{C_{1}(t)D_{1}(t)+C_{2}(t)D_{2}(t)},
\label{equ21}
\end{eqnarray}
while the components of electric field read:
\begin{eqnarray}
\tilde{E}_{xx}(t)=E_{x}[D_{3}(t)\tilde{\omega}_{cy}(t)+C_{3}(t)\lambda_{\pi_{x}}(t)+\dot{C}_{3}(t)],\nonumber\\
\tilde{E}_{xy}(t)=E_{x}[C_{3}(t)\tilde{\omega}_{cx}(t)-D_{3}(t)\lambda_{\pi_{y}}(t)-\dot{D}_{3}(t)].
\label{equ22}
\end{eqnarray}
As seen,
the dynamics is governed by the non-stationary coefficients.
It should be noted that the cross component $\tilde{E}_{xy}(t)$ of electric field is absent at the initial time
and appears during the evolution of system.

\section{Linear coupling in coordinate with heat bath}
\subsection{Solution of Non-Markovian Langevin equations}
For the system with linear coupling in coordinate, the coupling term is written as
\begin{eqnarray}
H_{cb}=\sum_{\nu}(\alpha_{\nu}x + \beta_{\nu}y)(b_\nu^+ + b_\nu)+
\sum_{\nu}\frac{1}{\hbar\omega_{\nu}}(\alpha_{\nu}x + \beta_{\nu}y)^{2},
\label{equ23}
\end{eqnarray}
where $\alpha_{\nu}$ and $\beta_{\nu}$ are real coupling
constants. Here, we again introduce the counter term which depends
on the coordinates of the collective system and can be treated as
a part of the potential. The operators of random forces and
dissipative kernels in Eqs. (\ref{equ8})   are
$$F_{x}(t)=-\sum_{\nu}{}\alpha_{\nu}(f_{\nu}^{+}+f_{\nu}),\hspace{.1in}
F_{y}(t)=-\sum_{\nu}{}\beta_{\nu}(f_{\nu}^{+}+f_{\nu})$$
and
\begin{eqnarray}
K_{xx}(t-\tau)&=& \sum_{\nu}^{}\frac{2\alpha_{\nu}^{2}}{\hbar\omega_\nu}\cos(\omega_\nu [t-\tau]), \nonumber\\
K_{yy}(t-\tau)&=&
\sum_{\nu}^{}\frac{2\beta_{\nu}^{2}}{\hbar\omega_\nu}\cos(\omega_\nu
[t-\tau]),
\label{equ24}
\end{eqnarray}
respectively.
Here, we assume that there are no correlations between $F_x^\nu$
and $F_y^\nu$, so that $K_{xy}=K_{yx}=0$.
If the coupling constants $\alpha_{\nu}$ and $\beta_\nu$ depend on  magnetic field, then
the dissipative kernels $K_{xx}$ and  $K_{yy}$
are the functions of $B$.

It is convenient to introduce the spectral density $D_{\omega}$
of the  heat bath excitations  to replace the sum
over different oscillators, $\nu$, by an integral over the
frequency: $\sum_{\nu}^{}...\to \int\limits_{0}^{\infty}d\omega
D_{\omega}...$. This replacement is accompanied by the following
replacements: $\alpha_\nu\to\alpha_{\omega}$, $\beta_\nu\to\beta_{\omega}$,
$\omega_\nu\to\omega$, and $n_\nu\to n_{\omega}$.
Let us consider the following spectral functions \cite{Katia}
\begin{eqnarray}
D_{\omega}\frac{|\alpha_{\omega}|^2}{\hbar\omega}=
\frac{\alpha^2}{\pi}\frac{\gamma^2}{\gamma^2+\omega^2},\hspace{.3in}
D_{\omega}\frac{|\beta_{\omega}|^2}{\hbar\omega}=
\frac{\beta^2}{\pi}\frac{\gamma^2}{\gamma^2+\omega^2},
\label{equ25}
\end{eqnarray}
where the memory time $\gamma^{-1}$ of the dissipation  is inverse
to the phonon bandwidth of the heat bath excitations which are
coupled to a quantum particle. This is  the Ohmic
dissipation with the Lorentian cutoff (Drude dissipation)
\cite{Kampen1,Kampen2,Kampen3,Kampen4,Kampen5,Kampen6,Katia,Kanokov}.

Using the spectral functions (\ref{equ25}), we obtain the dissipative kernels and their
Laplace   transforms in convenient forms
\begin{eqnarray}
K_{xx}(t)&=&m_x \lambda_x\gamma e^{-\gamma|t|},\hspace{.3in}
K_{yy}(t)=m_y \lambda_y\gamma e^{-\gamma|t|},\nonumber\\
K_{xx}(s)&=&\frac{m_x \lambda_x\gamma}{s+\gamma},\hspace{.6in}
K_{yy}(s)=\frac{m_y \lambda_y\gamma}{s+\gamma},
\label{equ26}
\end{eqnarray}
where the
coefficients
$$\lambda_x=\hbar \alpha^2=\frac{1}{m_x}\int_0^\infty K_{xx}(t-\tau)d \tau, \hspace{.3in}
\lambda_y=\hbar \beta^2=\frac{1}{m_y}\int_0^\infty K_{yy}(t-\tau)d
\tau$$
 are the friction
coefficients in Markovian limit.
So,  the solutions for the collective
variables (\ref{equ17}) include the following time-dependent
coefficients:
\begin{eqnarray}
A_{1}(t)&=&\dot{A}_{3}(t),\hspace{.3in}
A_{2}(t)=\dot{B}_{3}(t)|_{x\leftrightarrow y},\nonumber\\
A_{3}(t)&=&\frac{1}{m_x}(\frac{\lambda_y}{\lambda_x\lambda_y+\omega_{c}^2}t+\frac{\omega_{c}^2(\gamma-\lambda_y)-\lambda_y^2(\gamma-\lambda_x)}{\gamma(\lambda_x\lambda_y+\omega_{cx}\omega_{cy})^2}\nonumber\\
&+&\sum_{i=1}^4 \frac{b_{i}e^{s_it}(\gamma+s_{i})(\gamma\lambda_y+s_{i}(\gamma+s_{i}))}{s_{i}^{2}}),\nonumber\\
B_{1}(t)&=&\dot{A}_3(t)|_{x\leftrightarrow y},\hspace{.3in}
B_2(t)=\dot{B}_{3}(t),\nonumber\\
B_{3}(t)&=&\frac{\omega_{cx}}{m_y}\left(\frac{t}{\lambda_x\lambda_y+\omega_{cx}\omega_{cy}}+\frac{2
\lambda_x\lambda_y-\gamma(\lambda_x+\lambda_y)}{\gamma(\lambda_x\lambda_y+\omega_{cx}\omega_{cy})^2}
+\sum_{i=1}^4 \frac{b_{i}e^{s_it}(\gamma+s_{i})^{2}}{s_{i}^{2}}\right),\nonumber\\
C_{1}(t)&=&m_x\ddot{A}_{3}(t),\hspace{.3in} C_{2}(t)=m_x
\ddot{B}_{3}(t),\hspace{.3in}
C_3(t)=m_x\dot{A}_3(t),\nonumber\\
D_1(t)&=&C_1(t)|_{x\leftrightarrow y},\hspace{.3in} D_2(t)=m_y
\ddot{B}_{3}(t),\hspace{.3in} D_3(t)=m_y\dot{B}_{3}(t),
\label{equ27}
\end{eqnarray}
where $b_i=[\prod_{j\neq i}(s_i-s_j)]^{-1}$ with $i,j=1,2,3,4$ and
$s_i$ are the roots of the  equation
\begin{eqnarray}
\gamma\lambda_x
[\gamma\lambda_y+s(\gamma+s)]+(\gamma+s)(s[s^{2}+\omega_{c}^2]
+\gamma[\omega_{c}^2+s(\lambda_y+s)])=0.
\label{equ28}
\end{eqnarray}

\subsection{Asymptotic friction coefficients, renormalized cyclotron frequency, and  components of the electric field}
Using the
relationship $s_1s_2s_3s_4=\gamma^2(\lambda_{x}\lambda_{y}+\omega_{c}^2)$ between the roots of Eq.~(\ref{equ28}),
  we
obtain the asymptotic ($t\rightarrow\infty$) expressions   for the friction coefficients
\begin{eqnarray}
\lambda_{\pi_{x}}(\infty)=-\frac{[\gamma+s_1+s_2]
[\gamma \lambda_x+\omega_{c}^2+(s_1+\gamma)(s_1+s_2)+s_2^2]}{(\gamma+s_1+s_2)^2+\omega_{c}^2},\nonumber\\
\lambda_{\pi_{y}}(\infty)=-\frac{[\gamma+s_1+s_2][\gamma
\lambda_y+\omega_{c}^2+(s_1+\gamma)(s_1+s_2)+s_2^2]}{(\gamma+s_1+s_2)^2+\omega_{c}^2},
\label{equ29}
\end{eqnarray}
  renormalized frequencies
\begin{eqnarray}
\tilde{\omega}_{cx}(\infty)=\frac{\omega_{cx}[(s_1+\gamma)(s_2+\gamma)-\gamma
\lambda_x]}{(\gamma+s_1+s_2)^2+\omega_{c}^2},\nonumber\\
\tilde{\omega}_{cy}(\infty)=\frac{\omega_{cy}[(s_1+\gamma)(s_2+\gamma)-\gamma
\lambda_y]}{(\gamma+s_1+s_2)^2+\omega_{c}^2},
\label{equ30}
\end{eqnarray}
  renormalized cyclotron frequency
\begin{eqnarray}
\tilde{\omega}_{c}(\infty)=\omega_c\frac{[(s_1+\gamma)(s_2+\gamma)-\gamma
\lambda_x]^{\frac{1}{2}}[(s_1+\gamma)(s_2+\gamma)-\gamma
\lambda_y]^{\frac{1}{2}}}{(\gamma+s_1+s_2)^2+\omega_{c}^2}=
\sqrt{\tilde{\omega}_{cx}\tilde{\omega}_{cy}},\hspace{.3in}
\label{equ31}
\end{eqnarray}
and   components of the electric field
\begin{eqnarray}
\tilde{E}_{xx}(\infty)&=&\frac{E_{x}}{\lambda_x\lambda_y+\omega_{c}^2}
[\omega_{cx}\tilde{\omega}_{cy}(\infty)+\lambda_y\lambda_{\pi_{x}}(\infty)],\nonumber\\
\tilde{E}_{xy}(\infty)&=&\frac{E_{x}}{\lambda_x\lambda_y+\omega_{c}^2}[\lambda_y
\tilde{\omega}_{cx}(\infty)-\omega_{cx}\lambda_{\pi_{y}}(\infty)],
\label{equ32}
\end{eqnarray}
where $s_1$ and $s_2$ are the roots with the smallest absolute
values of their real parts.
 In the case of zero external
magnetic field ($B=0$), or Markovian limit
($\gamma\rightarrow\infty$), the cross current disappears because
  of $\tilde{E}_{xy}(\infty)=0$. Moreover it also
disappears at $\lambda_y=0$. Thus, if the particle
can move freely in the cross direction, i.e.  its
time-of-flight in this direction is
$\tau_y\sim1/\lambda_y=\infty$, there is no cross electric field
($E_{xy}=0$). This important result follows from Eqs. (\ref{equ29})--(\ref{equ32}).
So, in the superconductive materials
the Hall phenomenon should not be observed.

\subsection{Axial symmetric system}
One can obtain  clearer physical picture of the process,
if the space-symmetric system is considered. In this system
$m_x=m_y=m$, $\lambda_x=\lambda_y=\lambda$, and
$\omega_{cx}=\omega_{cy}=\omega_c$. So, the equation
(\ref{equ28}), which defines the poles in the integrands of $I_j$ and $I'_j$ ($j=x,y,\pi_x,\pi_y$), is
simplified:
\begin{eqnarray}
(s^2+\omega_c^2)(\gamma+s)^2+2\gamma \lambda s
(\gamma+s)+\lambda^2 \gamma^2=0.
\label{equ33}
\end{eqnarray}
This equation has analytic
roots:
\begin{eqnarray}
s_1&=&-\frac{1}{2}\left(\gamma+i \omega_c+\sqrt{(\gamma-i
\omega_c)^2-4 \gamma \lambda}\right),\hspace{.3in}
s_2=s_1^*,\nonumber\\
s_3&=&-\frac{1}{2}\left(\gamma+i \omega_c-\sqrt{(\gamma-i
\omega_c)^2-4 \gamma \lambda}\right),\hspace{.3in}
s_4=s_3^*.\nonumber
\end{eqnarray}
In order to split the real and imaginary parts of the roots, we
expand them up to the first order in $\lambda/\gamma$:
\begin{eqnarray}
s_1&=&-\frac{\lambda
\gamma^2}{\gamma^2+\omega_c^2}-i\frac{\omega_c^2+\gamma^2+\lambda
\gamma}{\gamma^2+\omega_c^2}\omega_c, \nonumber\\
s_3&=&-\gamma\frac{\gamma^2+\omega_c^2-\gamma
\lambda}{\gamma^2+\omega_c^2}+i\frac{\lambda \gamma
\omega_c}{\gamma^2+\omega_c^2}.\nonumber
\end{eqnarray}
Using the expansion of
\begin{eqnarray}
\tilde{\omega}_c&=&\tilde{\omega}_c(\infty)=\frac{\omega_c}{2}-
\frac{i}{4}\left(\sqrt{(\gamma-i \omega_c)^2-4 \gamma \lambda}+
\sqrt{(\gamma+i \omega_c)^2-4 \gamma \lambda}\right),\nonumber\\
\lambda_\pi&=&\lambda_{\pi_x}(\infty)=\lambda_{\pi_y}(\infty)=\frac{\gamma}{2}+\frac{i}{4}\left(\sqrt{(\gamma-i
\omega_c)^2-4 \gamma \lambda}-\sqrt{(\gamma+i \omega_c)^2-4 \gamma
\lambda}\right)
\label{equ34}
\end{eqnarray}
 up to the
first order in $\lambda/\gamma$
\begin{eqnarray}
\tilde{\omega}_c&=&\omega_c(1+\lambda_\pi/\gamma), \nonumber\\
\lambda_\pi &=&\frac{\gamma^2}{\gamma^2+\omega_c^2}\lambda,
\label{equ35}
\end{eqnarray}
we obtain from Eq. (\ref{equ32})
the analytical expressions for the components of electric field:
\begin{eqnarray}
\frac{\tilde{E}_{xx}(\infty)}{E_x}&=&1+\frac{\lambda_\pi}{\gamma},\nonumber\\
\frac{\tilde{E}_{xy}(\infty)}{E_x}&=&=\frac{\omega_c\lambda}{\gamma^2+\omega_c^2}=\frac{\omega_c\lambda_\pi}{\gamma^2}=
\frac{{\tilde{\omega}_c}-\omega_c}{\gamma}.
\label{equ36}
\end{eqnarray}
As  seen,  $\tilde{E}_{xy}(\infty)\to 0$  at $\gamma\to \infty$ or at $\lambda\to 0$, or $\omega_c\to 0$.
From the expression  for the
non-diagonal component $\tilde{E}_{xy}(\infty)/E_x$ of electric field, we can find the magnitude of
magnetic field when it reaches the maximum:
\begin{eqnarray}
\omega_c^{max}\approx\gamma.
\label{equ37}
\end{eqnarray}
The formula (\ref{equ37})
may serve as the simplest way of definition of the
memory time $\gamma^{-1}$ of the dissipation in the system.
Thus, in order to determine $\gamma$ of any system, it is
necessary to determine the magnetic field at which the resulting
cross electric field  $\tilde{E}_{xy}(\infty)$   reaches its maximum value.

\section{Calculated results and discussions}
In the  model  considered, one can
study
friction,  cyclotron frequency, and   external parameters of the system:
the longitudal and cross components of   electrical field.
In the calculations we set $\lambda=\lambda_x=\lambda_y$ (or $\lambda_\pi=\lambda_{{\pi}_x}=\lambda_{{\pi}_y}$)
and
$m=m_x=m_y$ (or $\omega_{c}=\omega_{cx}=\omega_{cy}$).
%

\subsection{Friction coefficient,  renormalized cyclotron frequency, and electrical field}
 The  dependencies of $\lambda_\pi$,
 $\tilde{\omega}_c$,
$\tilde{E}_{xx}$, and $\tilde{E}_{xy}$ on   time
are shown in Figs.~1 and 2.
The non-Markovian correction to the friction coefficient
increases with asymptotic friction coefficient (right side) and
decreases with the magnetic field (left side). The
increasing friction and magnetic field contribute  to the
rise of asymptotical microscopic magnetic field (bottom parts of
Fig.~1).   One can see in Fig.~2 that the cross electric field
increases with the magnetic field, while the correction to the
longitudinal electric field decreases. In general, the rise of the
asymptotic friction coefficient increases the transient time of
$\lambda_{\pi}, {\tilde\omega}_c, \tilde{E}_{xx}$, and
$\tilde{E}_{xy}$.   The change rate  of the cyclotron frequency is about $({\tilde\omega}_c-\omega_c) \gamma/2$.
One should reveal the reason of an increase of the energy of  cyclotron rotation.
The  magnetic forces  are perpendicular to the
velocity of charge carrier and do not affect the energy.
The dissipation and external magnetic field  affect each other due to the
non-Markovian dynamics of   quantum system and the value of
 magnetic field is changed.

The asymptotic behaviors of the transport coefficients
considered above are shown in   Figs.~3 and 4.
The effective friction
coefficient  decreases with increasing value of
$\omega_c$ (Fig.~3).
Note  that the  resistance obtained
in our model  does not depend on the magnetic field
because we neglect the influence of magnetic field on the
coupling between the quantum particle and heat-bath (or on the dissipative kernels).
 Moreover the curves
corresponding to larger $\gamma$ have a weaker decreasing tendency
  and the friction does not depend on   magnetic
field in the Markovian limit, $\gamma\rightarrow\infty$. In the
plot showing the dependence of effective friction
$\lambda_\pi$ on the Markovian friction $\lambda$, the line is inclined
less then 45 degrees to the abscissa. This means the friction
coefficient has relatively small influence   the system
  excepting the cases of very weak magnetic fields.

Analyzing the dependence
of the frequency of microscopic magnetic field $\tilde{\omega}_c$
(Fig.~3) and effective electric  field $\tilde{E}_{xx}$  (Fig.~4) on $\lambda$,
one can  conclude  that the  specimen with nonzero friction is more susceptible to the external magnetic
and electrical fields.
The non-Markovian corrections to  the external
magnetic  and electrical fields are larger
for the system with larger time of response $\gamma^{-1}$. In
the upper parts of Fig.~4, the behavior  of the cross
electrical field   is demonstrated. Relating the plots in Fig. 4,
the dependence of the tangent of the Hall angle $\tilde{E}_{xy}(\infty)/\tilde{E}_{xx}(\infty)$
 on friction and
frequency of external magnetic field are obtained, and the key conclusions might be
made. Firstly, the classical Hall phenomenon could not be observed at zero
magnetic field or in the systems with zero friction coefficient.
A
number of experiments  with superconductors \cite{Lewis}
supports this conclusion.
Secondly, according to the  curves given in   Fig.~4, the tangent of the Hall
angle reaches its maximum at the strength $\omega_c\approx\gamma$ of magnetic field.
Thirdly,  Eqs.~(35)
and our numerical calculations suggest that the cross
electric field is not originated in the Markovian limit ($\gamma\to\infty$).
Thus, taking non-Markovian nature of the system into account, one can
  explain the   Hall phenomenon.


\subsection{Application of  model to interpretation of  Hall angle experiment}
To demonstrate the possibilities of the model, we calculate  the
tangent of the Hall angle,
$\tan[\Theta_H]=\tilde{E}_{xy}(\infty)/\tilde{E}_{xx}(\infty)$, for the sample of
Zn settled in the increasing external magnetic field at two
temperatures. Many experiments were performed to measure this value
in several elements. We choose Zn because it has one type of charge carriers~\cite{Borovik} and
consequently easy to understand the technique of implementation of
the model. In the case of matters with more then one type of charge
carriers the problem is more complicated, since the two-band model
given above should be considered. In order to turn to the
 observable values in expressions (\ref{equ32}), all parameters in
these expressions should be multiplied by the  mass to charge ratio
$m/e$.
As a result, instead of the friction coefficient $\lambda$,
cyclotron frequency $\omega_c$, and inverse response time $\gamma$ of the system
 we  have the inverse reciprocal mobility of charge carriers
1/$\mu$, intensity of the magnetic field $B$ and new parameter
$\Gamma=m \gamma/e$. From the experimental data~\cite{Borovik} we
may define the strength of the magnetic field at which the charge
carriers deviate to the maximal angle from their non-field
direction. Knowing the field, we define the parameter $\Gamma$
by using expression (\ref{equ37}). Calculated and experimental
characteristics of Zn are given in Table I. The calculations performed with
values of mobility given in the table are in a good agreement with the
experimental data (Fig.~5), specially at high strength of magnetic field.


\section{Summary}
The behavior of the  generated flow of charge carriers under the influence of
external magnetic
and electric fields was investigated in the two-dimensional
case using the non-Markovian Langevin approach
and the general coupling between charge carriers and environment.
   The developed model was applied to the case where the collective
    variables are linear coupled in the coordinate with the variables of the
heat-bath. In order to average the influence of  heat-bath on the
collective system, we applied the spectral function
of   heat-bath excitations which describes Drude dissipation with Lorenzian cutoffs.
The classical Hall effect  was considered.
We showed that
the cross
electric field (which is absent at the initial time) does not appear in the Markovian limit.
So, taking non-Markovian nature of the system into consideration, one can
explain the classical Hall effect.
The dependence of the tangent of the Hall angle on the magnetic field
was investigated. Its
 value   increases up to a specific magnitude and  then monotonically decreases.
  The position of the maximum is defined by the memory time $\gamma^{-1}$.
One can suggest the method for determining memory time by measuring
the magnetic field at which the resulting cross electric field has a maximum value.
The possibilities of the model were shown through its
application to interpret the experiment for the Zn sample. The
calculated results agree well  with the experimental data.

\acknowledgments
This work was partially supported by  the Russian Foundation for Basic Research (Moscow)   and DFG (Bonn).
The IN2P3(France)-JINR(Dubna) Cooperation
Programme is gratefully acknowledged.


\begin{table}[h!]
    \caption{Experimental (asterisks)~\cite{Borovik} and theoretical
    characteristics of Zn at two temperatures.}\label{1}
    \begin{flushleft}
  \begin{tabular*}{\textwidth}{@{\extracolsep{\fill}}       c   c  c  c  c}
    \hline
     Temperature$^\ast$,& Resistance$^\ast$, $\rho_{xx},$& Mobility, & $\Gamma=B_{max}^{\ast},$ &Max. Hall  \\
      $T$  (K)& $\times 10^{-11}$ (Ohm $\cdot$ m) &$\mu$ (m$^2$ / V$\cdot$ s )  & (Tesla) & angle$^\ast$, $\Theta_H$    \\ \hline
      4.22& 2.6555 & 50.25 & 0.37 & $1.6\,^{\circ}$   \\
      20.4 & 35.595& 1.1 & 1.35 & $9.37\,^{\circ}$  \\
    \hline
   \end{tabular*}
  \end{flushleft}
\end{table}

\begin{figure}[h]
    \includegraphics[width=15cm]{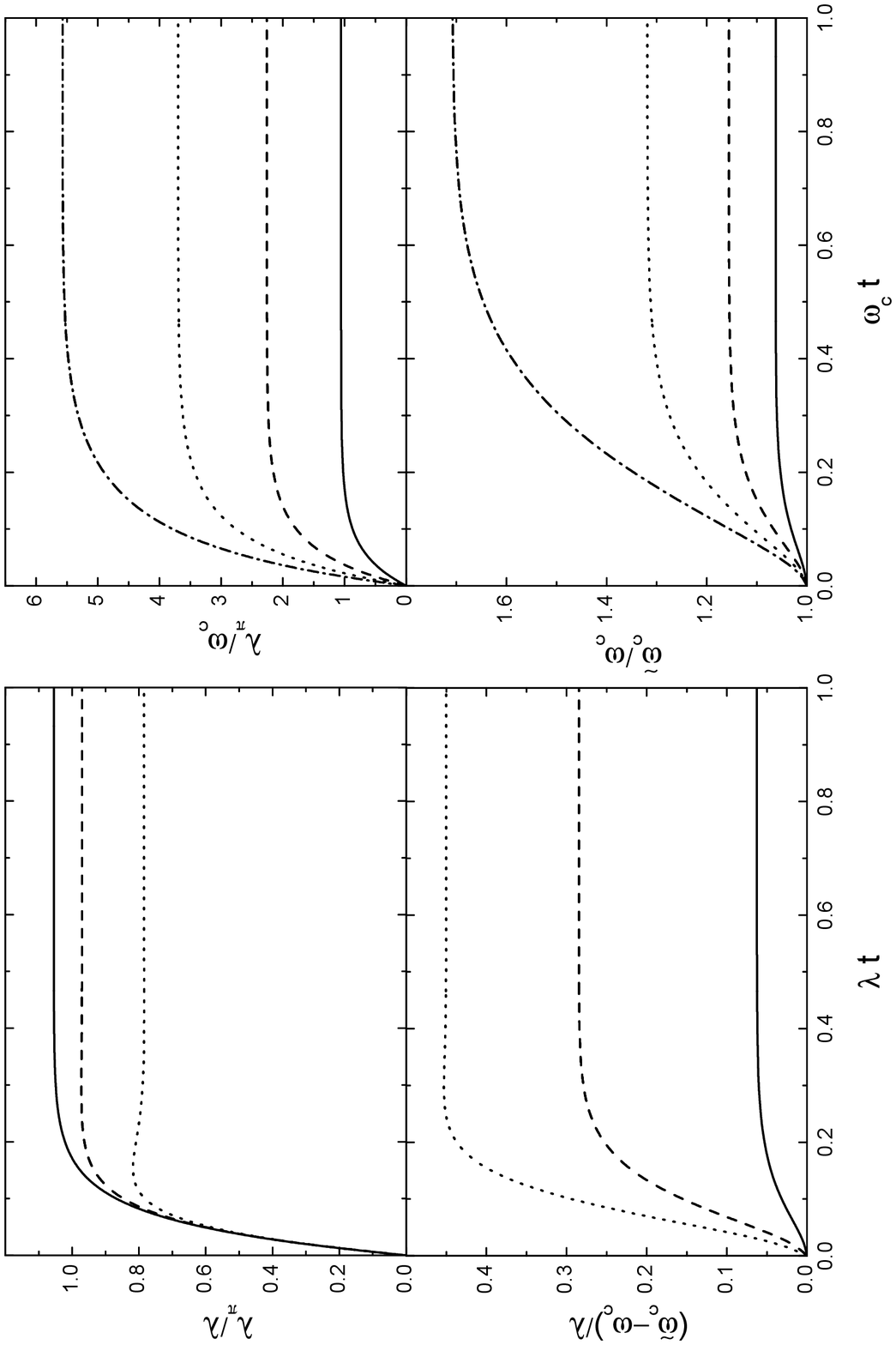}
\caption{The calculated  friction coefficient and cyclotron frequency as  functions of time.
  The results for the  frequencies $\frac{\omega_c}{\lambda}$=1, 5, and 10  of the external magnetic
 field   at the fixed Markovian friction coefficient $\lambda$
 are presented by solid, dashed, and dotted lines, respectively (left side).
 The results for the  Markovian friction coefficient  ($\lambda=\lambda_x=\lambda_y$)
 $\frac{\lambda}{\omega_c}$=1, 2,  3, and
4  at the fixed external magnetic
 field $\omega_c$ are presented by solid, dashed, dotted, and dash-dotted lines,
respectively (right side).}
\label{1_fig}
\end{figure}

\begin{figure}[h]
    \includegraphics[width=15cm]{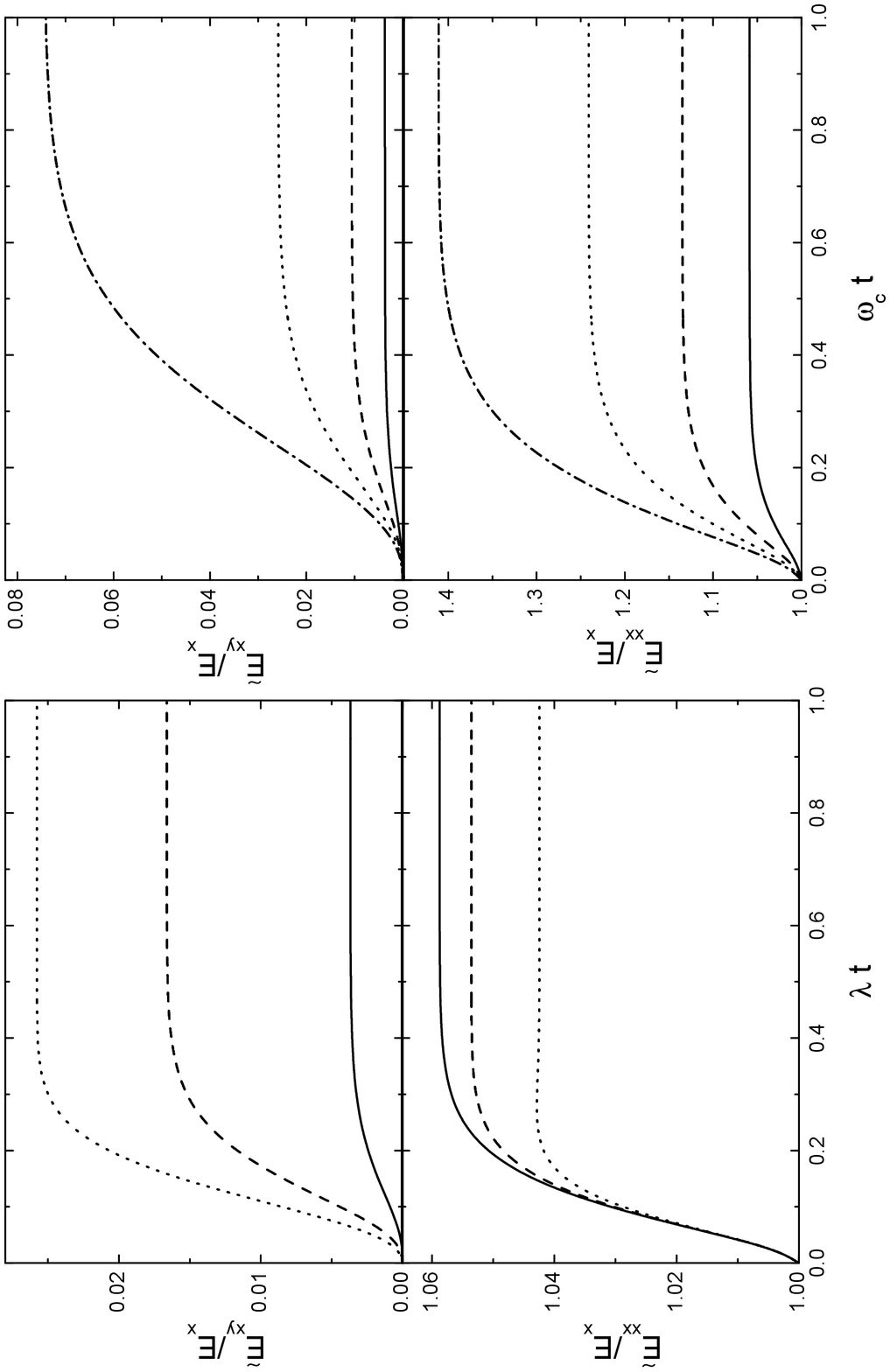}
\caption{The same as in Fig.~1, but for  the components of  electric field.}
\label{2_fig}
\end{figure}

\begin{figure}[h]
    \includegraphics[width=15cm]{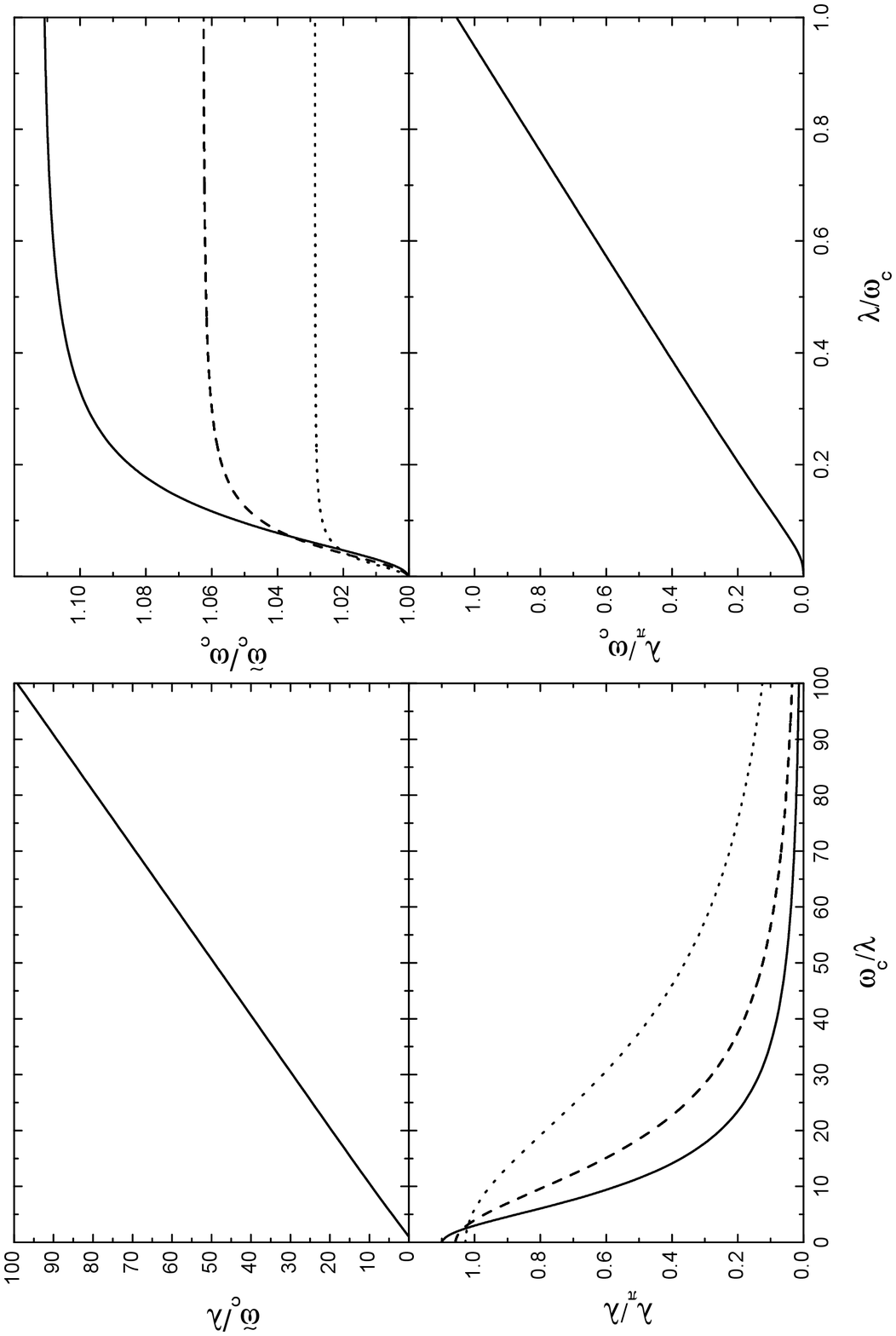}
  \caption{The calculated asymptotic cyclotron frequencies
   and friction coefficients as functions of $\omega_c$  (left side) and $\lambda$
   (right side). The solid, dashed, and dotted lines correspond to the calculations with $\gamma/\lambda$=12, 19, and
 38, respectively.}
\label{3_fig}
\end{figure}

\begin{figure}[h]
    \includegraphics[width=15cm]{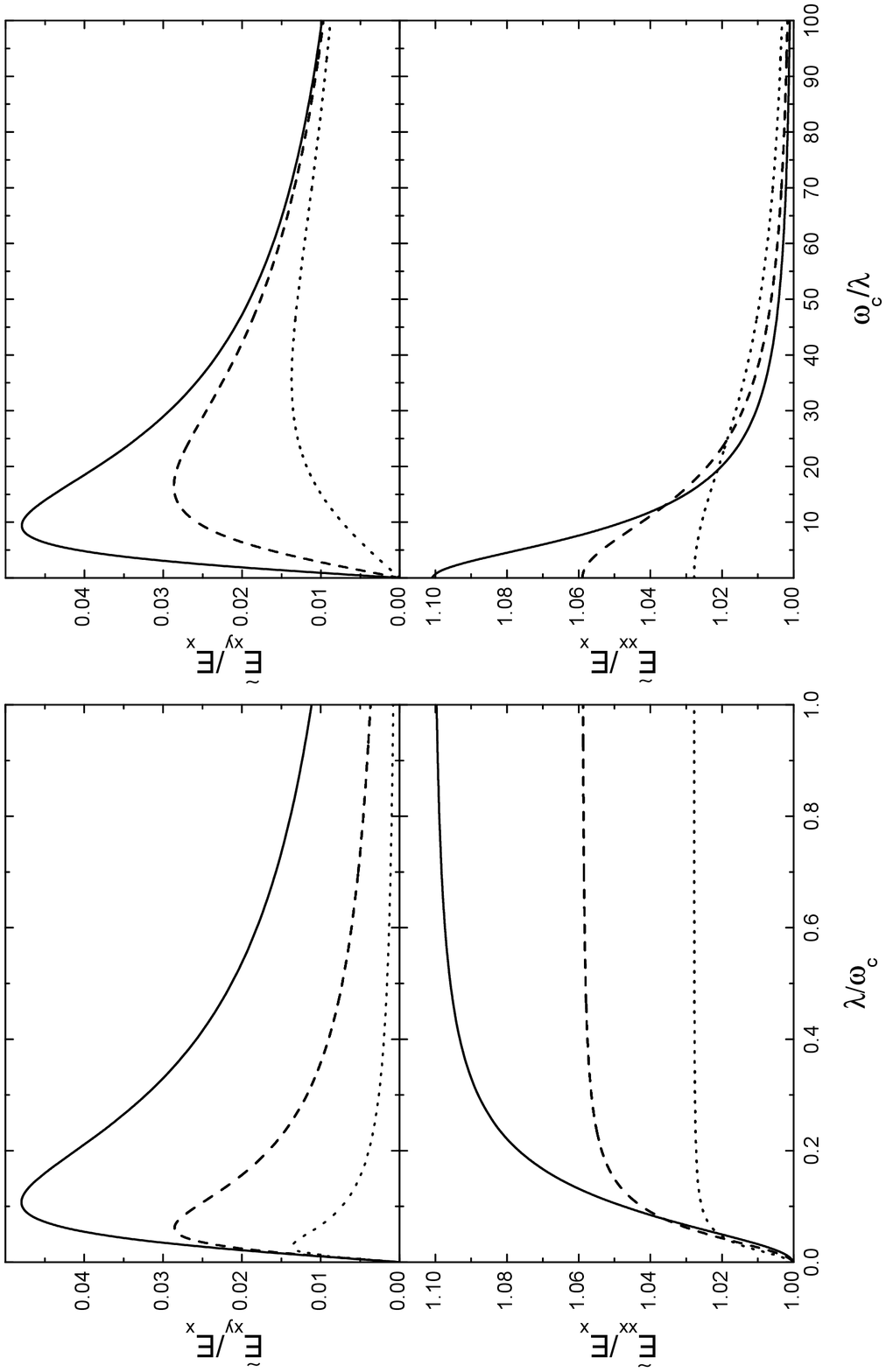}
  \caption{The same as in Fig.~3, but for the
   calculated asymptotic components of electric field.}
\label{4_fig}
\end{figure}

\begin{figure}[h]
    \includegraphics[width=15cm]{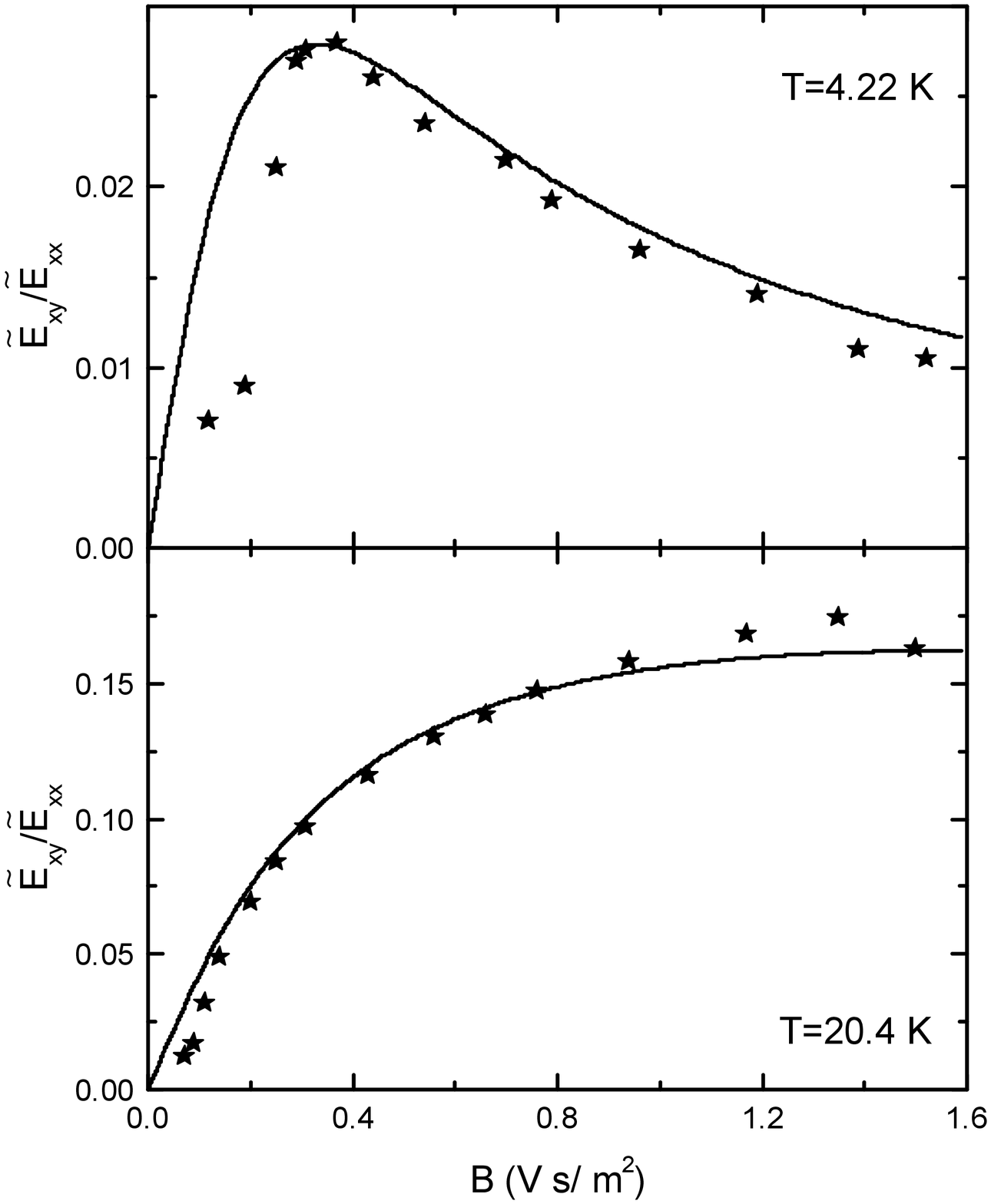}
\caption{The experimental~\cite{Borovik} (symbols) and theoretical dependencies (lines) of the
    tangent of the Hall angle on magnetic field, $B$,
    for zinc at the   temperatures $T$ indicated.}
\label{5_fig}
\end{figure}

\end{document}